\documentstyle [12pt]{article}
\title{
\hspace{3.0truein}{\small IFT-498-UNC}\\
\vspace{0.2truein}
{Fractional Exclusion Statistics and Anyons}
}
\author{Wei Chen\footnotemark[1]~$^1$
and Y. Jack Ng\footnotemark[2]~$^{1,2}$ \\
$^1$ Institute of Field Physics, Department of Physics
and Astronomy\\
University of North Carolina,
Chapel Hill, NC 27599-3255\\
and \\
$^2$ School of Natural Sciences\\ Institute for Advanced Study,
Princeton, NJ 08540}
\date{
}
\footnotetext{$^*$ chen@physics.unc.edu}
\footnotetext{$^\dagger$  ng@physics.unc.edu}

\begin{document}
\maketitle
\vspace{0.2truein}
\begin{abstract}
Do anyons, dynamically realized by the field theoretic
Chern-Simons construction, obey fractional exclusion statistics?
We find that they do if the statistical interaction between
anyons and anti-anyons is taken into account.
For this anyon model, we show perturbatively
that the exchange statistical parameter
of anyons is equal to the exclusion statistical parameter. We obtain the
same result by applying the relation between the exclusion statistical
parameter and the second virial coefficient in the non-relativistic limit.
\end{abstract}
\newpage
\baselineskip=22.0truept

The concept of {\it anyons} \cite{Me}-\cite{Wi1} has been useful in the
study of two spatial dimensional systems, particularly in the theory of
fractional quantum Hall effect \cite{La}, and possibly in the theory
of high $T_c$ superconductivity \cite{Wi2}. Anyons are particles whose
wave-functions acquire an arbitrary phase $e^{i(1-\alpha)\pi}$
when two of them are braided. They obey fractional statistics, with the
two limiting cases $\alpha = 0, 1$ corresponding to fermions and
bosons respectively. Several years ago Haldane \cite{Hal} introduced another
definition of fractional statistics based on the so-called {\it generalized
Pauli exclusion principle}. Haldane defined the statistical interactions
$g_{ij}$ for a system of $N$ particles of various species by
$g_{ij} = -\Delta d_i/\Delta N_j$, where $N_j$ is the number of
particles of species $j$, and $d_i$ is the dimension of the
single particle Hilbert space of the $i$-th species, obtained by
holding the coordinates and the species of $N-1$ particles fixed.
Conventional bosons have $g_{ij}=0$ while the Pauli exclusion principle for
fermions corresponds to $g_{ij} = \delta_{ij}$. Unlike the anyon fractional
{\it exchange} statistics, Haldane's fractional
{\it exclusion} statistics is formulated in arbitrary spatial
dimensions.

Recently there has been much interest in the
physics of the latter fractional statistics \cite{MS}-\cite{NW}.
In particular,
by examing the high temperature limit of the second virial coefficient,
the authors of Ref.\cite{MS} have shown that anyons satisfy Haldane's
new definition of fractional statistics, when generalized to
infinite dimensional Hilbert spaces.
However, even if the virial expansion is valid, their treatment is less
than complete. A correct description of matter at high temperatures
must take into
account the possibility of pair creation, but these authors obviously have
not considered the
effects of anti-particles. In this Letter, we reexamine the
issue whether anyons obey fractional exclusion statistics in the framework of
quantum field theory. We will use the model of anyons dynamically
realized by the Chern-Simons construction.

In Euclidean space ($g_{\mu\nu}=\delta_{\mu\nu}$) the model is given by
\begin{equation}
{\cal L} =  \psi^\dagger
[\gamma_\mu(\partial_\mu - ia_\mu) + iM]\psi
- \frac{i}{4\pi\alpha}
\epsilon_{\mu\nu\lambda}a_\mu\partial_\nu a_\lambda\;,
\label{L}
\end{equation}
where $\psi$ is a
two-component spinor, and $a_\mu$ is the Chern-Simons field.
The normalization of the Chern-Simons term is so chosen that
when $\alpha =0$ and $1$, this theory describes free fermions
and free (hard core) bosons, respectively \cite{PCI}.
When $0<\alpha<1$, this is a theory for ``free'' anyons, as the
Chern-Simons field carries no local dynamical degree of freedom and its
sole role is to endow the charged particles with ``magnetic'' flux tubes,
via the minimal interaction.
Fixing the sign of the mass term, we choose $\alpha > 0$ so that the
fluxes attached to the particles
are in the positive $z$ direction.

The symmetries of the theory have been discussed
in \cite{DJT}.  Under a $U(1)$ gauge transformation,
Eq.(\ref{L}) changes by a total derivative.
Accordingly, this theory has a $U(1)$
global symmetry with a conserved charge (particle number)
$N=\int d^2x \psi^\dagger\gamma_0\psi\;.$
The charge conjugation (C) transformation leaves the lagrangian
invariant, while the fermion mass and the Chern-Simons terms are both
variant under parity (P) or time reversal (T). But CPT symmetry
holds.

At finite temperature, the functional integral representation of the
partition function for the Chern-Simons-fermion system given by
Eq.(\ref{L}) is
\begin{equation}
Z =\int{\cal D}a_\mu
{\cal D}{\psi}^\dagger
{\cal D}\psi{\cal D}\bar{c}{\cal D}c
exp\left[-\int_0^\beta d\tau\int d^2x {\cal L}_1\right]\;,
\label{Z}
\end{equation}
with
\begin{equation}
{\cal L}_1=
{\psi}^\dagger[\gamma_\mu(\partial_\mu -ia_\mu)+ iM
+ \mu\gamma_0]\psi
-\frac{i}{4\pi\alpha}\epsilon_{\mu\nu\lambda}a_\mu\partial_\nu a_\lambda
 +\frac{1}{2\rho}(\partial_\mu a_\mu)^2
+ (\partial_\mu\bar{c})(\partial_\mu c)\;,
\label{Se}
\end{equation}
where $\beta = 1/T$ ($k_B=1$) is the inverse of the
temperature;
a chemical potential $\mu$ is introduced, to ensure
particle number conservation.
As is clear from Eq.(\ref{Se}), the ghost field $c$ does not
interact with any other fields; it serves to
cancel the non-physical gauge field degrees of freedom.
In the present case, the path integral over the ghost fields
cancels
exactly the Gaussian integration over the Chern-Simons kinetic term.
At finite temperature, boson (fermion) field is
subject to periodic (anti-periodic)
boundary condition,
$a_\mu(\beta, {\bf x})=a_\mu(0, {\bf x})\;,$
and $\psi(\beta,{\bf x}) = - \psi(0,{\bf x})\;.$
Accordingly, the imaginary time $\tau$ is
mapped to discontinuous frequencies $\omega_n$
in the momentum space,
$\omega_n = 2\pi n T$ for bosons,
and $(2n+1)\pi T$ for fermions.

The Feynman rules for the model Eq.(\ref{Se})
in the Landau gauge ($\rho = \infty$) are
given by ${\cal S}_0 = 1/i(\gamma^\mu p_\mu + M)\;,
{}~ p_0 = (2n+1)\pi T + i\mu\;;$
${\cal D}^{\mu\nu}_0 =
-2\pi\alpha\epsilon^{\mu\nu\lambda}p_\lambda/p^2\;,~
p_0 = 2n\pi T\;;$ and
$\Gamma_0^\mu =i\gamma^\mu\;.$

We consider perturbation expansion around the free theory, assuming
$\alpha$ small. To the leading order in $\alpha$, without the interaction
term, the system describes an 
ideal fermion gas.
Performing the Gaussian integrations in Eq.(\ref{Z}), we readily obtain
\begin{equation}
{\rm ln}Z_0 = V\int\frac{d^2{\bf p}}{(2\pi)^2}\left[\beta\omega
+ ln(1+e^{-\beta\omega}z)+ln(1+\frac{e^{-\beta\omega}}{z})\right]\;,
\label{lnZ0}
\end{equation}
where $\omega = \sqrt{{\bf p}^2+M^2}$ is
 the single particle energy
and  $z= e^{\beta\mu}$ the fugacity;
 the contribution
from the integral over Chern-Simons term is canceled by that from
the ghost \cite{Ch}. The first term in Eq.(\ref{lnZ0}) is
the zero-point energy, and the last two terms are contributions
from the fermions and anti-fermions.

At the next leading order, we consider the two-loop vacuum diagram

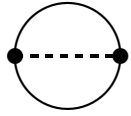
\unitlength=1.00mm
\linethickness{0.4pt}
\thicklines
\begin{picture}(110.0,38.0)
\put(58.00,18.00){\circle*{2.00}}
\put(72.00,18.00){\circle*{2.00}}
\put(65.00,18.00){\circle{40.00}}
 \multiput(58.0,18.0)(2.00,0.00){7}{\line(3,0){1.00}}
\end{picture}

\noindent The Feynman integral of the diagram is
\begin{eqnarray}
& &-2\pi\alpha T^2\sum_n\int\frac{d^2{\bf q}}{(2\pi)^2}
\sum_m\int\frac{d^2{\bf p}}{(2\pi)^2}
{\rm tr}\left(\gamma_\mu\frac{1}{i(\gamma_\lambda p_\lambda + M)}
\gamma_\nu\frac{1}{i(\gamma_\rho q_\rho + M)}\right)
\frac{\epsilon_{\mu\nu\sigma} (q-p)_\sigma}{(q-p)^2}\nonumber\\
&=& -8\pi\alpha M\left(
T\sum_n\int\frac{d^2{\bf q}}{(2\pi)^2}\frac{1}{q^2_0+{\bf q}^2+M^2}\right)^2\;.
\end{eqnarray}
It is easy to check that the summation takes the form
\begin{equation}
T\sum_n\frac{1}{q^2_0+\omega^2}
= \frac{1}{2\omega}\left(1
- \frac{1}{e^{\beta\omega}/z+1}
 - \frac{1}{e^{\beta\omega}z+1}\right)\;,
\label{sum}
\end{equation}
with $\omega=\sqrt{{\bf q}^2+M^2}$.
Now consider the integration over the
two dimensional momentum ${\bf q}$.
The integral for the first term is linearly divergent,
therefore regularization and
renormalization are needed. Since this term is temperature independent,
we choose to renormalize it away
so that the free energy is zero at $T=0$.
The integral for the rest of Eq.(\ref{sum}) is finite.
Then we obtain the first order correction
\begin{equation}
{\rm ln}Z_1=  \frac{\alpha MTV}{4\pi}\left(
{\rm ln}(1+ \frac{e^{-\beta M}}{z})
+{\rm ln}(1+ e^{-\beta M}z)\right)^2 \;.
\label{lnZ1}
\end{equation}
We see that with the Chern-Simons interaction, the particles
and anti-particles (now anyons and anti-anyons) interfere, even
in the ``free'' anyon theory. Given ${\rm ln}Z$, one can determine
all thermodynamical functions. For instance, the pressure
of the anyon gas is $P = ({\rm ln}Z_0+{\rm ln}Z_1+...)/\beta V$.
{}From Eq.(\ref{lnZ1}), we see that attaching the charged fermions
with flux tubes in the positive $z$ direction increases the
pressure of the system.
If the flux tubes attached to the fermions are in $-z$ direction,
the pressure is decreased. This is a reflection of
Parity variance of the anyon system.
This seems to suggest a possible
means to test the nature of the anyons by experiments (if any).

The particle density is $n= N/V = \beta z\partial P/\partial z$. To
first order in $\alpha$, we obtain the anyon density
$n=n^{(0)}+n^{(1)}$, with
\begin{eqnarray}
n^{(0)} &=&
\frac{MT}{2\pi}\left(
{\rm ln}(1+ e^{-\beta M}z)
-{\rm ln}(1+ \frac{e^{-\beta M}}{z})\right.\nonumber\\
& & \left.~~~~~~~~~~~-\frac{1}{\beta M}\left[
{\rm dilog}(1+e^{-\beta M}z)-
{\rm dilog}(1+\frac{e^{-\beta M}}{z})\right]
\right)\;,
\label{n0}\\
n^{(1)} &=& \frac{\alpha MT}{2\pi}\left(
{\rm ln}(1+ e^{-\beta M}z)
+{\rm ln}(1+ \frac{e^{-\beta M}}{z})\right)
\left(\frac{1}{1+ e^{\beta M}/z}
-\frac{1}{1+ e^{\beta M}z}\right)
\label{n2}\;.
\end{eqnarray}

We now turn to discuss particle densities for systems obeying fractional
exclusion statistics. We start with Wu's formulation \cite{Wu}
of quantum statistical
mechanics in terms of Haldane's statistical interactions.
Let $n_1=n_1(\epsilon)$ and $n_2=n_2(\epsilon)$ denote the
distribution function of the 1st and 2nd
 species of particles
with single particle energy $\epsilon_1$ and $\epsilon_2$,
 chemical potential $\mu_1$ and $\mu_2$,
and exclusion statistics $g_1$ and $g_2$, respectively. If
the statistical interactions between the two species
(called mutual statistics in Ref.\cite{Wu})
are $g_{12}$ and $g_{21}$, then
\begin{equation}
n_1e^{\beta(\epsilon_1-\mu_1)}=
(1+g_1n_1-g_{12}n_2)^{g_1}(1-n_1+g_1n_1-g_{12}n_2)^{1-g_1}
\left(\frac{1-g_{21}n_1-n_2+g_2n_2}{1+g_2n_2-g_{21}n_1}\right)^{g_{21}}\;,
\end{equation}
and another equation given by interchanging $1$ and $2$.
In the above, $g_i=0, 1$ ($i=1$ and $2$) correspond to fermions and bosons,
respectively. (Our notations are related to Haldane's by $g_{ii}=1-g_i$.)
Expanding
the above equation to first order in
$g_1$, $g_2$, $g_{12}$ and $g_{21}$, we obtain
\begin{equation}
n_1=
n_1^0+g_1n_1^0\left(n_1^0-(1-n_1^0){\rm ln}(1-n_1^0)\right)-g_{12}n_1^0n_2^0
+g_{21}n_1^0(1-n_1^0)
{\rm ln}(1-n_2^0) + ... \label{n1}
\end{equation}
and a similar expression given by interchanging $1$ and $2$. We have used
the notation $n^0_i = 1/(e^{\beta(\epsilon_i-\mu_i)}+1)$ for the free
fermion distribution.

For the dynamical system under consideration, the two species of
particles correspond to the particles and anti-particles;
by charge conjugation invariance,
$g_1=g_2=-g_{12}=-g_{21}=g$.
The particle density is $\bar{n}=n_1-n_2$, with
chemical potential $\mu=\mu_1=-\mu_2$, and single particle
(anti-particle) energy, $\epsilon=\epsilon_1=\epsilon_2$.
The distribution function,  to the next leading
order linear in $g$, is
$\bar{n}(\epsilon) = \bar{n}^{(0)}(\epsilon) + \bar{n}^{(1)}(\epsilon)$
with
\begin{eqnarray}
\bar{n}^{(0)}(\epsilon) &=& \frac{1}{e^{\beta\epsilon}/z+1}
-\frac{1}{e^{\beta\epsilon}z+1}\;,\label{dis0}\\
\bar{n}^{(1)}(\epsilon) &=&
g\left(
[{\rm ln}(1+e^{-\beta\epsilon}z)
+{\rm ln}(1+e^{-\beta\epsilon}/z)]
[\frac{e^{\beta\epsilon}/z}{(e^{\beta\epsilon}/z+1)^2}
-\frac{e^{\beta\epsilon}z}{(e^{\beta\epsilon}z+1)^2}]
\right.\nonumber\\
& &~~~+\left.\frac{1}{(e^{\beta\epsilon}/z+1)^2}
-\frac{1}{(e^{\beta\epsilon}z+1)^2}\right)
\label{dis}\;.
\end{eqnarray}

If we use $\epsilon = {\bf p}^2/(2M) + M$
as the single particle energy
in Eq.(\ref{dis}), the integration over ${\bf p}$
yields a particle density
with exactly the same expression given in Eq.(\ref{n2}) obtained from
the anyon theory, provided that we identify
\begin{equation}
\alpha = g\;.
\label{id}
\end{equation}
This implies that, as far as the particle density
is concerned, the anyons obey the exclusion
statistics. In particular, the interference of anyons
and anti-anyons is related to the statistical interaction
between
different particle species. To establish this connection,
the non-relativistic, large $\beta M$, limit  has been taken
in the lowest order $n^{(0)}$ of the anyon theory
(so that the dilog terms in Eq.(\ref{n0}) can be ignored).

It is tempting to compare the expressions of
the distribution functions. The one
for the anyon model to first order in $\alpha$ can be calculated
by using Eq.(\ref{lnZ0}) and Eq.(\ref{lnZ1}). The result is
$n({\bf p}) = n^{(0)}({\bf p})+n^{(1)}({\bf p})$ with
\begin{eqnarray}
n^{(0)}({\bf p}) &=& \frac{1}{e^{\beta\omega}/z+1}
-\frac{1}{e^{\beta\omega}z+1}\;,\label{n00}\\
n^{(1)}({\bf p})&=&\frac{\alpha}{2}
\frac{M}{\omega}\left(
[{\rm ln}(1+e^{-\beta M}z)
+{\rm ln}(1+\frac{e^{-\beta M}}{z})]
[\frac{e^{\beta\omega}/z}{(e^{\beta\omega}/z+1)^2}
-\frac{e^{\beta\omega}z}{(e^{\beta\omega}z+1)^2}]
\right.\nonumber\\
& &~~~~~~~~+\left.[\frac{1}{e^{\beta M}/z+1}
-\frac{1}{e^{\beta M}z+1}][\frac{1}{e^{\beta\omega}/z+1}
+\frac{1}{e^{\beta\omega}z+1}]\right)\;.
\label{n22}
\end{eqnarray}
Comparing these with Eq.(\ref{dis0}) and Eq.(\ref{dis}),
we find that only for ${\bf p}=0$
(so that $\epsilon = \omega = M$), the distribution function of anyons
takes exactly the same form as the one obtained from the analysis of
fractional exclusion statistics; even here there is a discrepancy,
viz $n^{(1)}$ in Eq.(\ref{n22}) is
half of $\bar{n}^{(1)}$ in Eq.(\ref{dis}),
if the anyon statistics is identified
with the exclusion statistics as in Eq.(\ref{id}).

Knowing the pressure and particle density for the anyon model, we can
derive the equation of state (and thereby the second virial coefficient).
We content ourselves with its non-relativistic (large $\beta M$) limit.
To the first order in $\alpha$, the particle density
$n=n^{(0)}+n^{(1)}$ for the anyon system is given by Eq.(\ref{n0})
and Eq.(\ref{n2}). In the large $\beta M$ limit we find \cite{note1}
\begin{equation}
\lambda^2 n \cong e^{-\beta M}(z-\frac{1}{z})
+ (\frac{1}{2}-\alpha) e^{-2\beta M}(\frac{1}{z^2}-z^2)\;,
\end{equation}
where 
$\lambda=\sqrt{2\pi\beta/M}$ is the thermal wavelength.
The fugacity and its reciprocal can
now be solved in terms of $\lambda^2 n$:
\begin{equation}
 z \cong \left\{ \begin{array}{c}
e^{\beta M}\lambda^2 n[1+(\frac{1}{2}-\alpha)]\lambda^2n]\\
                0\end{array} \right. \; ~~~{\rm and}~~~
\frac{1}{z} \cong \left\{ \begin{array}{c} 0\\
e^{\beta M}(-\lambda^2 n)[1-(\frac{1}{2}-\alpha)]\lambda^2n]
                \end{array} \right.
\label{z}
\end{equation}
for positive and negative $n$ respectively. We have assumed that
$\lambda^2 n \ll 1$, a necessary condition
for virial expansion which we will use in the following. On the other hand,
the pressure, given by $P \cong 
({\rm ln}Z_0+{\rm ln}Z_1)/\beta V$,
can be
read off from Eq.(\ref{lnZ0}) and Eq.(\ref{lnZ1}). In the non-relativistic
(large $\beta M$) limit, we readily get 
\begin{equation}
\beta P = \frac{e^{-\beta M}}{\lambda^2}(z+\frac{1}{z})
+ (-\frac{1}{4}+\frac{\alpha}{2})\frac{e^{-2\beta M}}
{\lambda^2}(z^2+\frac{1}{z^2}) + ...\;.
\label{P}
\end{equation}
Substituting Eq.(\ref{z}) into Eq.(\ref{P}) we obtain the
 equation of state for the anyon-anti-anyon system in the form of virial
expansion,
\begin{equation}
\beta P = |n|\left(1+(\frac{1}{4}
-\frac{\alpha}{2})\lambda^2 |n| + ...\right)\;.
\end{equation}
Thus the second virial coefficient is given by \cite{note2}
\begin{equation}
B_2 = \frac{1}{4}-\frac{\alpha}{2}\;.
\label{B2}
\end{equation}
Note that this expression is exact for both the fermion ($\alpha=0$)
and boson ($\alpha=1$) gases. It is tempting to speculate that
the result for anyon gases may be exact although we have obtained it
by perturbation expansion in $\alpha$ only.
Eq.(\ref{z}) indicates that in the large $\beta M$ limit, the $z$ term,
{\it i.e.} the anyon contribution (the $1/z$ term, {\it i.e.}
the anti-anyon contribution) dominates for the case of positive
(negative) particle density. Thus
we are now in a position to apply the relation, found recently by
Murthy and Shankar \cite{MS}, between the exclusion statistical
parameter and the second virial coefficient,
which, in our notation, reads
\begin{equation}
-\frac{1}{2}+g = -2B_2(\alpha)\:.
\end{equation}
(Here it is the large $\beta M$ approximation that validates Murthy and
Shankar's approach.)
The substitution of Eq.(\ref{B2}) into the above equation immediately
yields $g = \alpha\;.$
This is the same result we derived in Eq.(\ref{id}).

We thank S. Adler, Y.M. Cho, Z.N.C. Ha, and Y.-S. Wu for
useful conversations. One of us (YJN) thanks the faculty at
IAS for hospitality.
This work was supported in part by
the  U.S. DOE grant No. DE-FG05-85ER-40219 and by the
Z. Smith Reynolds fund of the University of North Carolina.


\begin{thebibliography}{199}
\bibitem{note1}
Actually there is also a term $e^{-\beta M}(z-1/z)/\beta M$ on
the right hand side of the equation.
But it can be shown that this term does not contribute to the
second virial coefficient in the large $\beta M$ limit.
\bibitem{note2}
This is to be compared with the result given by Ref.\cite{BHR}
for a first-quantized system of spin-$1/2$ flux-carrying fermions,
and the coefficient given by Ref.\cite{ASWZ}
for the corresponding spinless system.
Obviously the second virial coefficient for the
anyon system is model dependent.
\end{thebibliography}
\end{document}